\documentclass{ws-procs9x6}
\usepackage{latexsym}
\usepackage{amssymb}
\usepackage{amsmath}
\usepackage{bm}
\usepackage{subfigure}
\usepackage{float}
\usepackage{indentfirst}
\usepackage{array}
\usepackage{xspace}
\usepackage[dvips]{epsfig}
\usepackage{psfrag}
\usepackage{amscd}
\usepackage{slashbox}
\usepackage{hhline}

\newcommand{\lbr}{\left(}
\newcommand{\rbr}{\right)}

\newcommand{\lb}{\label}
\newcommand{\om}{\omega}

\newcommand{\ga}{\gamma}

\newcommand{\fr}{\frac}

\newcommand{\bw}{\begin{widetext}}
\newcommand{\ew}{\end{widetext}}
\newcommand{\be}{\begin{equation}}
\newcommand{\ee}{\end{equation}}
\newcommand{\ba}{\begin{eqnarray}}
\newcommand{\ea}{\end{eqnarray}}

\def\vr{{\bf r}}

\def\v{{\bf v}}

\def\e{{\rm e}}
\def\m{{\mu}}

\newcommand{\p}{\mathrm{p}}

\renewcommand{\leq}{\leqslant}

\begin{document}
\title{Electromagnetic and gravitational radiation from massless particles}

\author{D.\,V.\,Gal'tsov$^*$}

\address{Moscow State University, Faculty of Physics, Department of
Theoretical Physics \\
 $^*$E-mail: galtsov@phys.msu.ru}

\begin{abstract}
We demonstrate that full description of both electromagnetic and
gravitational radiation from massless particles lies outside the
scope of classical theory. Synchrotron radiation from the
hypothetical massless charge in quantum electrodynamics in external
magnetic field has finite total power while the corresponding
classical formula diverges in the massless limit. We argue that in
both cases classical theory describes correctly only the
low-frequency part of the spectra, while the total power diverges
because of absence of the UV frequency cutoff.  Failure of
description of gravitational radiation from massless particles by
classical General Relativity may be considered as another appeal for
quantization of gravity apart from the problem of singularities.
\end{abstract}

\bodymatter

\medskip
The limit of zero mass of radiating charge in classical
electrodynamics is non-trivial. As was discussed recently
\cite{Azzurli:2012tp,Lechner:2014kua}, the Lienard-Wiechert
potentials, due to the factor $(1-\vr\v/r)$ in the denominator,
 diverge in the direction of the instantaneous velocity for $|\v|=1$.
Attempts to regularize this singularity had not led to reasonable
results, so it was argued in \cite{Lechner:2014kua} that the truth
is that the massless charge does not radiate at all. Similar
conjecture  was promoted before on different grounds
\cite{Xiang:2007gw}. This, however, apparently contradicts to
infiniteness of the massless limit $\m\to 0$ of the well-known
formula for synchrotron radiation  \cite{Sokolov1}:
 \be\lb{clas}
P_{\rm cl}=\frac{2e^4 H^2}{3\m^2}\left(\frac{E}{\m}\right)^2\,,
 \ee
where $E$ denotes the energy of the charge, and $H$ -- the magnetic
field. This discrepancy led us to investigate this problem in the
quantum theory \cite{Gal'tsov:2015cla}.

Radiation from massless charges in quantum electrodynamics is also a
non-trivial problem.  Within the perturbation theory one encounters,
apart from the usual infrared divergencies, the {\em collinear
singularities}, occurring when the photon is emitted from the
massless legs of  the Feynmann diagrams in the direction of the
momentum of a charge \cite{Weinberg:1965nx}. This is manifestation
of degeneracy of states of the  charge and the photon moving along
the same line. Elimination of collinear divergences is achieved
using Kinoshita-Lee-Nauenberg \cite{Kinoshita:1962ur,Lee:1964is}
prescription of averaging over an ensemble of degenerate states.
Note that the mentioned above line singularities of  classical
retarded potentials look similar to collinear singularities of
quantum theory. Another complication is the screening of the
electromagnetic field of the massless charge  due to vacuum
polarization \cite{Vaks:1961}. However, these problems are evaded,
if interaction of the charge with an external electromagnetic field
is treated non-perturbatively, using the exact operators in the
external classical field. In the case of magnetic field one deals
with the bound states of the charge at Landau levels, and the
momentum in the plane orthogonal to magnetic field is not conserved.
This removes collinear divergences and modifies the propagation
function, leading, in particular, to  non-zero quantum correction to
mass in the one-loop order. In this approach radiation from the
massless charge is non-zero and finite \cite{Gal'tsov:2015cla}.

 This analysis also reveals that
classical theory still can be applied if only one computes not the
Lienard-Wiechert potentials, but their spectral expansion. Indeed,
in view of the correspondence principle, it can be expected that
classical theory describes correctly the low frequency part of
radiation whatever the mass of the radiating charge is. In the
massive theory \cite{Sokolov1} one has for the low frequencies:
 \be \lb{dP} \frac{dP}{d\omega}=\frac{e^2
\omega_{H}3^{1/6}\Gamma(2/3)}{ \pi }
\left(\frac{\omega}{\omega_H}\right)^{1/3} \,,
  \ee
where $\omega_H=eH/E$. This expression depends only on the particle
energy and it is unchanged for zero mass. The difference between
massive and massless particles, however, is that in the former case
the formation length $l\sim 1/(\ga\omega_H)$ of radiation in a given
direction is finite, leading to the frequency cutoff at $\omega_{\rm
cr}\sim l^{-1} \gamma^2\sim\omega_H\ga^3$, where $\gamma$ is the
Lorentz-factor $\gamma=E/\mu=1/\sqrt{1-v^2}$. In the massless case
$\omega_{\rm cr}\to\infty$, so there is no frequency cutoff. The
Lienard-Wiechert potentials in coordinate representation account for
the total field, so they are inappropriate for description of
radiation from the massless charge indeed. But performing the
spectral decomposition, one can give a reasonable estimate using
classical theory with the quantum cutoff $\omega_{\rm
quant}=E/\hbar$. Integrating (\ref{dP}) up to this cutoff one
obtains:
 \be\lb{cut}
 P_{\rm cut}=\int_0^{\omega_{\rm quant}}\frac{dP}{d\omega}d\omega=
 \frac{e^2 \sqrt{3} \Gamma(2/3)}{4\pi \hbar^2} (3e\hbar HE)^{2/3}
   \,.
  \ee

Transition to the massless limit in the quantum theory of
synchrotron radiation of massive charge
\cite{Sokolov1,Schwinger:1973} is subtle, since the results of the
latter depend on two dimensionless parameters (in the units
$\hbar=c=1$):
 \be\lb{param}
f=\frac{H}{H_0}=\frac{eH}{\m^2}\,,\qquad \chi= \frac{H}{H_0}
\frac{E}{\m}=\frac{eHE}{\m^3}\,,
  \ee
 diverging as $\m\to 0$. Because of this, the standard approximations
used in the quasiclassical case (high Landau levels in both initial
and final sates) in terms of the Macdonald or Airy functions fail,
and one has to develop an alternative approximation scheme. We used
\cite{Gal'tsov:2015cla} the mass-operator in the Schwinger formalism
\cite{Schwinger:1973,Tsai:1974cc} (see also \cite{Baier:1990qq})
setting the charge mass to zero  {\em ab initio} and obtaining the
spectral power as the integral
 \be
\frac{dP}{d\omega}= \frac{e^2v}{4\pi E}\int_0^\infty
  \left(E^2(8-v^2)(1-v)^2x\sin\psi\;
 +\frac{eHv}{x^2}(1-\cos\psi)\right)dx\,,
 \ee
where  $v=\omega/E$ and  $\psi =\frac{x^3 E^2}{3eH} v(1-v)^2$.
Evaluating the integral over $x$  we get
 \be\lb{speless}
\frac{dP}{d\omega}=\frac{2e^2\;\Gamma\left(2/3\right)}{27\hbar
E}\,\left(3e\hbar HE\right)^{2/3}{\cal P}\left(
{\hbar\omega}/{E}\right)\,,
 \ee
 where the Planck's constant is restored, and the
normalized spectral function is introduced \be \lb{spenorm} {\cal
P}\left(v\right)=\frac{27}{2\pi\sqrt{3}}\;v^{1/3}(1-v)^{2/3}\,,\qquad
\int_0^1{\cal P}\left(v\right)dv=1\,.
 \ee
This spectrum is  smooth, exhibiting maximum  at \be \hbar
\omega_{\rm max}=\frac13 E\,. \ee The average   photon energy is
 \be\lb{ave}
\langle\hbar\omega\rangle=E\int_0^1  v{\cal P}\left(v\right)dv=
\frac49 E \,,
 \ee
while the total energy loss per unit time reads
 \be\lb{tot}
P=\int_0^{E/\hbar} P(\omega)d\omega=
\frac{2e^2\;\Gamma\left(2/3\right)}{27\hbar^2 }\;\left(3e\hbar
HE\right)^{2/3}\,.
 \ee
It differs from the estimate (\ref{cut}) only by a numerical
coefficient. The expression (\ref{speless}) has the following
unusual features. It is non-perturbative in the fine structure
constant $e^2/hc$, and it has no classical limit $\hbar\to 0$, being
essentially quantum. This could be expected in view of the Eq.
(\ref{ave}).

Now we are going to show that gravitational radiation form massless
particles exhibits similar features, though, as could be expected
from the Weinberg's theorems on soft photons and gravitons
\cite{Weinberg:1965nx}, the corresponding divergencies are weaker
(logarithmic). Note that, while consideration of massless charges in
electrodynamics appeals rather to the limiting case of the realistic
theory, in gravity it is not so, since all particles, including
truly massless ones (photons) are subject to gravitational
interaction and thus are entitled to emit gravitational radiation.
In the textbooks it is tacitly assumed that General Relativity
describes correctly gravitational radiation from any classical
sources. This turns out not to be true. Namely, the spectrum of
gravitational radiation from massless particles moving along null
geodesics in curved space-time, being computed classically, is UV
divergent. This can be considered as the second argument (apart from
the problem of classical singularities) appealing to quantization of
gravity.

Gravitational radiation from  massive bodies moving along
ultrarelativistic geodesics around black holes (gravitational
synchrotron radiation, GSR) was considered in early 1970-ies  as
possible mechanism of enhancement of the flux of gravitational waves
from the center of Galaxy to explain the (unconfirmed) Weber's
results. Calculations were performed in the Schwarzschild
\cite{Misner:1972jf,Khriplovich:1973bq} and Kerr
\cite{Chrzanowski:1974nr,Gal'tsov :1987ht} geometries.  Here we
discuss the massless limit in the GSR theory, which is interesting
not only in view of the above conceptual problem, but also in view
of discovery of high energy astrophysical sources involving strong
fluxes of photons and neutrinos.

Consider for simplicity the Schwarzschild case. Timelike geodesics
parameterized by the proper time $\tau$ obey the  radial equation $
 ({dr}/{d\tau})^2+U(r)=0$
with the effective potential $$U(r)=\lbr 1-\frac{2M}{r}\rbr \lbr
\fr{L^2}{r^2}+1\rbr-\ga^2\,,$$ where $\ga=E/\mu$ as before and $L$
is the angular momentum. The most interesting are the winding orbits
which perform many turns around black hole before being scattered
back, or absorbed by  the hole. To estimate gravitational radiation
one considers  circular orbits, whose radii $r_\p$ correspond to two
conditions $U(r_\p)=0=U'(r_\p)$. Solving these equations with
respect to $\gamma$ and $L$, one finds $
 \ga= \lbr 1- {2M}/{r_\p}\rbr  \lbr
 1- {3M}/{r_\p}\rbr^{-1/2}\,,\quad  {L}/{\ga}=(Mr_\p)^{1/2}\lbr
1- {2M}/{r}\rbr^{-1}\,, $ while the rotation frequency  in terms of
the orbit radius $r_\p$ reads $ \om_0= {d\phi}/{dt}=\lbr
{M}/{r_\p}\rbr^{1/2}$. Relativistic time-like circular orbits with
$3M<r_\p<6M$ are unstable. Those which lie in the interval
$3M<r_\p<4M$ become unbound under small perturbations, they
correspond  to large angle scattering with the impact parameter $ b=
{L}/{\lbr\ga^2-1\rbr^{1/2}}=r_\p\lbr
  {4M}/{r_\p}-1\rbr^{-1/2}
$
In the ultrarelativistic case $\ga\gg 1$ the unbound orbits with the
impact parameter close to the critical value
$
 b=3\sqrt{3} M
$ scatter  on the black hole with multiple revolutions, radiation
from these orbits can be estimated using the GSR power. The limit
$\ga=\infty$ corresponds to the massless particles (photon orbits),
the corresponding rotation radius being $ r_{\rm ph}=3M\,.
 $
Its dislocation with respect to timelike ultrarelativistic orbits
can be characterized by a small parameter $\delta$ via $
r_\p=(3+\delta)M $. Two useful relations then follow in the leading
order in $\delta$: $  {dt}/{d\tau}=\sqrt{ {3}/{\delta}}\,,\; \ga^2=
1/{3\delta}. $

Radiation field is expanded in terms of the appropriate angular
harmonics, labeled by two integers $l,\;m,\;|m|\leq l$, the main
contribution for large $\ga$ coming from $|m|\gg 1$ and $l$
differing from $|m|$ by $0,\;1$, depending on polarization. The
total intensity can be presented as a sum
 \be\lb{gsum}
 P_{\rm GSR}=\sum_{m=1}^\infty \frac{E^2\,\omega_0 }{M}\;F_m (r_p,
M)\;\frac{m_{\rm cr}}{m}\;\e^{-m/m_{\rm cr}}\,,
 \ee
where each term corresponds  to radiation with the frequency
$\omega=m\omega_0,\;F_m$ is a smooth function of the parameters, and
the critical frequency is
 \be\lb{gcut}
m_{\rm cr}=\fr{12}{\pi}\ga^2\,.
  \ee
The spectrum is therefore a falling function of the harmonic number
$m$, and it is cut off classically at the frequency $\ga$ times
smaller than in the case of flat space synchrotron radiation
(because of the increasing formation length due to closeness of
ultrarelativistic timelike geodesics of the radiating particle and
the null geodesics followed by gravitons \cite{Khriplovich:1973bq}).

The main contribution ($96\%$) to the total power comes from the
polarization commonly denoted as $\otimes$. This quantity was
computed in \cite{Gal'tsov :1987ht} (in earlier papers only the
distribution over the harmonics was given) and reads: \be P_{\rm
GSR}
=\fr{6\e^{-\pi/4}(r_\p-M)\left|\Gamma(1/4+i/4)\right|^2}{\pi^{3/2}r_\p^2(r_\p+3M)}
E^2 \ln(E/\mu)\,.
  \ee
The last factor reflects the logarithmic divergence of the sum
(\ref{gsum}) if the frequency cutoff (\ref{gcut}) is removed, what
happens in the limit of zero mass $\ga=\infty$. We therefore
conclude that General Relativity fails to provide  full description
of gravitational radiation form massless particles. Quantum theory
of gravitational synchrotron radiation is not developed yet, but one
can hope to get a correct estimate of the total power replacing the
divergent logarithm by $\ln(E/\hbar\omega_0)$ as we tested in the
electromagnetic case.

Radiation of scalar $s=0$ and vector $s=1$ waves is described
similarly to Eq. (\ref{gsum}) with different $m$-dependence, namely
$ ({m}/{m_{\rm cr}})^{1-s}$, with $s=2$ standing for gravitational
waves. Thus  the divergence of the total power as $\mu\to 0$ is
$\ga^2$ for $s=0,\,1$. Softer divergence in the gravitational case
is due to the fact that the effective gravitational coupling is
proportional to the energy, as was noted by Weinberg long ago.
\cite{Weinberg:1965nx}.

This work was supported by RFBR  under the project 14-02-01092-a.


\begin{thebibliography}{10}

\bibitem{Azzurli:2012tp}
  F.~Azzurli and K.~Lechner,
  Phys.\ Lett.\ A {\bf 377}, 1025 (2013)
  [arXiv:1212.3532];
  F.~Azzurli and K.~Lechner,
  Annals Phys.\  {\bf 349}, 1 (2014)
  [arXiv:1401.5721 [hep-th]].

\bibitem{Lechner:2014kua}
  K.~Lechner,
  J.\ Math.\ Phys.\  {\bf 56}, no. 2, 022901 (2015)
  [arXiv:1405.4805 [hep-th]].

\bibitem{Xiang:2007gw}
  L.~Xiang and Y.~G.~Shen,
  Int.\ J.\ Theor.\ Phys.\  {\bf 46}, 576 (2007);
  B.~P.~Kosyakov,
  J.\ Phys.\ A {\bf 41}, 465401 (2008)
  [arXiv:0705.1228 [hep-th]];
  Y.~Yaremko,
  Electron.\ J.\ Theor.\ Phys.\  {\bf 9}, no. 26, 153 (2012).

\bibitem{Sokolov1}
   A.~A.~Sokolov and I.~M.~Ternov, Synchrotron Radiation, Pergamon, New
   York, 1968).

\bibitem{Gal'tsov:2015cla}
  D.~V.~Gal'tsov,
  Phys.\ Lett.\ B {\bf 747}, 400 (2015)
  [arXiv:1505.06775 [hep-th]].

\bibitem{Weinberg:1965nx}
  S.~Weinberg,
  Phys.\ Rev.\  {\bf 140}, B516 (1965).

\bibitem{Kinoshita:1962ur}
  T.~Kinoshita,
  J.\ Math.\ Phys.\  {\bf 3}, 650 (1962).

\bibitem{Lee:1964is}
  T.~D.~Lee and M.~Nauenberg,
  Phys.\ Rev.\  {\bf 133}, B1549 (1964).

\bibitem{Vaks:1961}
  V.~G.~Vaks,
  Soviet\ Physics\ JETP  {\bf 13} 556-561 (1961);
  V.~N.~Gribov,
  Nucl.\ Phys.\ B {\bf 206}, 103 (1982).


\bibitem{Schwinger:1973}
   J.~Schwinger, Particles, Sources, and Fields (Addison-
   Wesley, Heading, Mass.), Vol. III, Chap. 5, Sec. 6;
   Phys. Rev. D 7, 1969 (1973).

\bibitem{Tsai:1974cc}
  W.~Y.~Tsai,
  Phys.\ Rev.\ D {\bf 8}, 3460 (1973).

\bibitem{Baier:1990qq}
  V.~N.~Baier, V.~M.~Katkov and V.~M.~Strakhovenko,
  Sov.\ Phys.\ JETP {\bf 71}, 657 (1990)
  [Zh.\ Eksp.\ Teor.\ Fiz.\  {\bf 98}, 1173 (1990)].

\bibitem{Misner:1972jf}
  C.~W.~Misner, R.~A.~Breuer, D.~R.~Brill, P.~L.~Chrzanowski, H.~G.~Hughes and C.~M.~Pereira,
  Phys.\ Rev.\ Lett.\  {\bf 28}, 998 (1972);
  R.~A.~Breuer, R.~Ruffini, J.~Tiomno and C.~V.~Vishveshwara,
  Phys.\ Rev.\ D {\bf 7}, 1002 (1973).

\bibitem{Khriplovich:1973bq}
  I.~B.~Khriplovich and E.~V.~Shuryak,
  Zh.\ Eksp.\ Teor.\ Fiz.\  {\bf 65}, 2137 (1973).

\bibitem{Chrzanowski:1974nr}
  P.~L.~Chrzanowski and C.~W.~Misner,
  Phys.\ Rev.\ D {\bf 10}, 1701 (1974).

\bibitem{Gal'tsov :1987ht}
  D.~V.~Gal'tsov  and A.~A.~Matyukhin,
  Yad.\ Fiz.\  {\bf 45}, 894 (1987).

\end{thebibliography}
\end{document}